# An active approach towards monitoring and enhancing drivers' capabilities - the ADAM cogtec solution.

Moti Salti, Yair Beery and Erez Aluf

ADAM cogtec inc.

Abstract

Driver's cognitive ability at a given moment is the most elusive variable in assessing driver's safety. In contrast to other physical conditions, such as short-sight, or manual disability cognitive ability is transient. Safety regulations attempt to reduce risk related to driver's cognitive ability by removing risk factors such as alcohol or drug consumption, forbidding secondary tasks such as texting, and urging drivers to take breaks when feeling tired. However, one cannot regulate all factors that affect driver's cognition, furthermore, the driver's momentary cognitive ability in most cases is covert even to driver.

Here, we introduce an active approach aiming at monitoring a specific cognitive process that is affected by all these forementioned causes and directly affects the driver's performance in the driving task. We lean on the scientific approach that was framed by Karl Friston (Friston, 2010). We developed a closed loop-method in which driver's ocular responses to visual probing were recorded. Machine-learning-algorithms were trained on ocular responses of vigilant condition and were able to detect decrease in capability due fatigue and substance abuse. Our results show that we manage to correctly classify subjects with impaired and unimpaired cognitive process regardless of the cause of impairment (77% accuracy, 5% false alarms).

Introduction

According to the National Highway Traffic Safety Administration (NHTSA), human error causes 94% of road accidents; in other words, most accidents are caused by humans' failure to correctly respond to the ever-changing situation on the road. Driving Monitoring Systems (DMS) technology are aimed at monitoring and alerting driver's engagement with the driving task but are often too late in preventing accidents and do not cover all causes for human errors. DMS relies on external driver behavior such as blinks and head nodding to diagnose one's ability to perform safely the driving task. However, this might prove to be too late. Closed eye lids and nodding might suggest that the driver is already asleep. These monitored behaviors are merely late symptoms of fatigue. Moreover, fatigue is not the only cause for driving performance decline. These causes vary from substance abuse to emotional distraction.

**Aim:** Our aim is to demonstrate a method for monitoring driver's ability to perform the driving task that (1) would be agnostic to cause of decline in abilities (2) would enable giving an early indication of driver's ability.

We introduce an active approach aiming at monitoring a specific cognitive process that is affected by all these forementioned causes and directly affects the driver's performance in the driving task. We lean on the scientific approach that was framed by Karl Friston (Friston, 2010). Friston's Predictive Coding approach treats the brain as a predicting machine, which constantly generates and updates a mental model of the environment. That model is used to generate predictions, which are then compared to actual sensory inputs; this comparison results in prediction errors, which are then used to update and revise the cognitive model. We capitalize on the fact that the efficiency of this 'prediction machine' would decrease in tandem with a decrease in vigilance (Bekinschtein et al., 2009), substance abuse or cognitive and emotional load (Tiferet-Dweck et al., 2016).

We developed a closed loop-method in which driver's ocular responses to visual probing were recorded. Machine-learning-algorithms were trained on ocular responses of vigilant condition and were able to detect decrease in capability due fatigue and substance abuse.

**Methods**

Participants: Total of 31 subjects were tested

Stimulus: stimulus was a changing train of LED lights lasting for 30 seconds. Stimulus was presented in subjects' visual periphery.

Procedure: three scenarios were tested not all subjects participated in all scenarios (see results). The first scenario was Alcohol use, the second was marijuana usage and third was fatigue. In the scenarios of substance abuse subjects were seated next to the driver and were instructed to

'command the driving mission'. In accordance with the driving literature, they were asked to make strategic and tactical driving decisions. Subjects participated in 10 shifts of 8 hours each. In 3 shifts they were asked to drink alcohol until their breathalyzer test showed 240 ug. In another 3 shifts they smoke a marijuana cigarette.

In the fatigue scenario subjects drove the car. Subjects participated in 10 shifts of 8 hours each. They used an adaptation of the KSS scale.

In all scenarios subjects' ocular and postural data was gathered with a standard DMS system and then 'recycled' offline to a another DMS system.

Classification:

Due to the assumption that we will initially have only sober data, we trained two types of classifiers for each subject: (a) a one-class classifier that was trained on one subject at a time on its sober data only and (b) a binary classifier that was trained on all subjects but one. We then created an ensemble of classifiers based on both types of classifiers where each classifier had the same weight. The one-class classifiers that we used were one-class SVM, isolation-forest and Local-Outlier-Factor (LOF). The binary classifier was XGBoost

Results

| Mobile | DMS 1 (native) | DMS 2 (cloud) | DMS 1 + DMS 2 |
|---|---|---|---|
| Overall | 36 subjects<br>77% Acc | 29 subjects<br>76% Acc | 29 subjects<br>83% Acc |
| Baseline (sober) | 36 subjects<br>5% FP | 29 subjects<br>20% FP | 29 subjects<br>7% FP |
| Alcohol | 21 subjects<br>83% Acc | 16 subjects<br>85% Acc | 18* subjects<br>85% Acc |
| THC | 13 subjects<br>64% Acc | 14 subjects<br>70% Acc | 11 subjects<br>80% Acc |
| Fatigue | 2 subjects<br>100% Acc | 2 subjects<br>100% Acc | 2 subjects<br>100% Acc |
| Declination due to long driving time (all data)** | 8 subjects<br>81% Acc | 6 subjects<br>75% Acc | 6 subjects<br>80% Acc |
| Declination due to long driving time (high amount of data)** | 5 subjects<br>100% Acc | 3 subjects<br>100% Acc | 3 subjects<br>100% Acc |

* Since we omit subjects that have same number of right and wrong classification checkups, we decrease the number of subjects. Combining the two DMSs, we break the tie thus having more subjects than DMS 2
** Binary classification

Discussion

Our results show that our method can reliably detect a decline in driver's subjective performance. Moreover, it is agnostic to the cause of this decline. Although our results were

Assessing a driver's predictive ability and cognitive state by recording the driver's behavior alone is particularly challenging (in behavior we refer to the various responses that could be recorded - including motor activity, physiological responses, ocular responses, etc.). Indeed, people exhibit a wide range of behaviors; but people, like other lifeforms, tend to present these behaviors in an adaptive manner. Accordingly, people do not perform the whole repertoire of their possible behaviors in a random manner; instead, they tend to have a restricted set of responses that correspond to a certain stimulus or a set of stimuli. Passively monitoring one's behavior to draw general conclusions about one's cognitive state in relation to one's environment would therefore be incredibly inefficient. A system that is only capable of observing behavior without equating the stimuli that had elicited that behavior cannot adequately interpret or predict the state of the driver.

**Conclusion:** Unlike other methods, we put forward an **active system**. By probing the driver with a known stimulus at a specific timing, to which we have already mapped the driver's repertoire of possible responses, we profoundly reduce the uncertainty of the system's predictive capabilities; we reduce that uncertainty even further by focusing on ocular responses. We chose to focus on ocular responses for several reasons: firstly, ocular responses were shown to be indicative of the driver's cognitive state at a specific moment (see our response to the previous concern), and secondly, ocular responses provide an early indication of a change in that state. DMS that attempt to prevent accidents by passively monitoring the driver's behavior (through indications such as eyelid status, head position, etc.) would simply be too late. If the driver's eyes are already closed and her/ his head is already drooping, then the DMS would be able to alert that the driver is asleep but would be able to do very little to prevent an accident. However, understanding the driver's actual cognitive state and trace a deterioration in that state would enable the system to deploy a hierarchy of intervention measures, such as notifying a fleet dispatcher, deploying an in-cabin alert, etc. Furthermore, employing an active approach enables us to establish a personalized, individual baseline for each driver; every driver, in every drive, would be evaluated according to her/his personal baseline.